\documentclass[10pt,journal,compsoc]{IEEEtran}

\usepackage{colortbl}
\usepackage{amsmath,amssymb,amsfonts}
\usepackage{algorithmic}
\usepackage{graphicx}
\usepackage{textcomp}
\usepackage{booktabs}
\usepackage{float}
\usepackage{multirow}
\usepackage{cite}
\usepackage{hyperref}

\hypersetup{
  colorlinks=true,       
  linkcolor=blue,        
  citecolor=blue,        
  urlcolor=blue          
}

\ifCLASSINFOpdf
\else
\fi
\begin{document}
\title{Context-Aware Service Recommendation System for the Social Internet of Things}

\author{
Amar Khelloufi, Huansheng Ning, Abdelkarim Ben Sada, Abdenacer Naouri and Sahraoui Dhelim.
\thanks{A. Khelloufi, H. Ning, A. Naouri, A. Ben Sada are with the School of Computer and Communication Engineering, University of Science and Technology Beijing, Beijing 10083, China}
\thanks{S. Dhelim is with the School of Computer Science, University College Dublin, Ireland, Corresponding author: Sahraoui Dhelim (email: sahraoui.dhelim@ucd.ie)}
\thanks{Manuscript received Month Day, Year; revised Month Day, Year.}
}

\maketitle

\begin{abstract}
The Social Internet of Things (SIoT) enables interconnected smart devices to share data and services, opening up opportunities for personalized service recommendations. However, existing research often overlooks crucial aspects that can enhance the accuracy and relevance of recommendations in the SIoT context. Specifically, existing techniques tend to consider the extraction of social relationships between devices and neglect the contextual presentation of service reviews. This study aims to address these gaps by exploring the contextual representation of each device-service pair. Firstly, we propose a latent features combination technique that can capture latent feature interactions, by aggregating the device-device relationships within the SIoT. Then, we leverage Factorization Machines to model higher-order feature interactions specific to each SIoT device-service pair to accomplish accurate rating prediction. Finally, we propose a service recommendation framework for SIoT based on review aggregation and feature learning processes. The experimental evaluation demonstrates the framework's effectiveness in improving service recommendation accuracy and relevance. 
\end{abstract}


\section{Introduction}
\label{sec:introduction}

The rapid growth and proliferation of interconnected devices and technologies have given rise to the Internet of Things (IoT), transforming our everyday lives in unprecedented ways. The IoT refers to the network of physical objects embedded with sensors, software, and connectivity, enabling them to collect and exchange data autonomously. However, as the IoT evolves, a new paradigm called the Social Internet of Things (SIoT) has emerged, extending the concept of connectivity beyond devices to include social interactions and human-centric services \cite{ASI}.
The SIoT represents a significant shift from a traditional device-centric model to a user-centric and socially interconnected environment. It utilizes the capabilities of IoT devices to not only enable communication but also establish social relationships between devices. By integrating social aspects into the IoT framework, the SIoT empowers devices to actively share information, collaborate, and engage in social interactions, leading to a revolutionary transformation in the discovery and composition of services \cite{zhang2022internet}. This paradigm shift facilitates the provision of personalized services, improves social interactions, and fosters innovative and meaningful connections among devices \cite{social_computing}.
With the rapid expansion of the SIoT ecosystem, the diversity of services and applications available has become overwhelming. SIoT entities find themselves inundated with a vast number of choices when it comes to selecting the right services. Consequently, there is a pressing need for effective service recommendation systems that can assist in identifying and suggesting the most relevant services for specific groups of users or devices \cite{contess}.

Several research works have focused on developing general service recommendation systems within the SIoT. However, these models often face significant challenges in adapting to the dynamic nature of the SIoT environment.  For instance, one proposed solution for analyzing the social correlation of service requirements is the Social Correlation Group-based Recommender System (SRS) introduced in \cite{kang2017srs}. SRS generates target groups based on the social correlation of service requirements, using an architecture and procedures derived from Collaborative Filtering and Content-based Recommender Systems. However, this social correlation service recommendation system has limitations. It primarily relies on profile similarity, friend similarity, and interest individuality, overlooking the diverse content and different modalities of data across devices and users. Additionally, the system fails to account for evolving user preferences and the integration of new data into the model, relying solely on historical user-item interactions. This limitation can result in inaccurate recommendations that do not align with users' current preferences or situations. Another proposal is the time-aware smart object recommendation model discussed in \cite{chen2019time}. This model considers users' preferences over time and the social similarity of objects to assist users in locating relevant smart objects in the SIoT. However, similar to previous works, this model lacks consideration for multi-modal data and contextual factors such as location and user characteristics. Incorporating these elements would enhance the accuracy and personalization of recommendations provided by the time-aware smart object recommendation model.
The graph-based service recommendation framework presented in \cite{chen2021graph} jointly considers social relationships between heterogeneous objects in the SIoT and user preferences. This framework models users, objects, and their relationships using a knowledge graph and learns users' preferences from their object usage events with a latent variable model. However, relying solely on user preferences to form the graph knowledge may result in a lack of accuracy and reliability in providing relevant services. Neglecting other types of data, such as contextual information, trends, or correlations of user-object relationships, may decrease the system's ability to tailor recommendations for individual users accurately.

In a different approach, \cite{khelloufi2020social} we proposed a service recommendation system that takes advantage of the relationships between device owners when recommending services. This system leverages effective community detection algorithms to identify specific communities and provide tailored service recommendations based on device owners' relationships within those communities. However, this work primarily focuses on device-device interactions and does not consider the service-service pair structure when dealing with large sparse data.

In previous research \cite{khelloufi2023multi}, we developed a service recommendation system that effectively utilizes multi-modal data, incorporates the diversity of generated data, and considers the relationships between different items/services within SIoT in order to get more accurate and personalized service recommendations to fit the dynamic and diverse SIoT environment. 

Therefore, Existing models in the SIoT space have not effectively leveraged contextual data to provide tailored recommendations. Moreover, these models have primarily focused on device-device and device-item relationships in service recommendation systems, neglecting the potential use of context-aware recommendations that can encompass review-based and interaction-based recommendations with user preferences and content heterogeneity. 

To overcome these limitations, we aim to develop a contextually aware service recommendation system that explores the contextual representation of device-service pairs and captures latent feature interaction to enhance the service recommendation system within the SIoT environment. Thus, mining the contextual data as well as device-service interaction within the SIoT is crucial for delivering tailored service recommendations that meet the diverse needs of SIoT objects, including users and devices.

The remainder of this article is organized as follows: Section \ref{sec:related_works} provides a comprehensive review of recent literature on service recommendations in the SIoT. In Section \ref{sec:system_design}, we present the system modeling, including the framework overview, the review aggregation process, and the engagement feature learning process. The experimental evaluation of our proposed system is discussed in Section \ref{sec:evaluation}. Finally, Section \ref{sec:Conclusion} concludes the paper, highlighting our research contributions, and outlines future research directions in the context-aware service recommendation systems for the SIoT.

\section{Related Works}
\label{sec:related_works}

The emergence of the SIoT has brought forth a multitude of interconnected devices, generating vast amounts of data and creating opportunities for novel services and applications. Within this context, service recommendation systems play a crucial role in facilitating personalized and efficient user experiences. This section explores the realm of service recommendation systems in the domain of SIoT, delving into existing research and approaches aimed at addressing the challenges of recommending services within a socially connected IoT environment. 
In our previous work \cite{khelloufi2020social}, we proposed a service recommendation system that capitalizes on the social relationships between device owners to enhance the accuracy and diversity of offered services in the IoT context. However, this system disregards potential limitations that arise from relying solely on users' social relationships, neglecting other crucial factors like user preferences, context, and specific requirements. Considering these additional aspects could further enhance the efficiency and personalization of the recommended services within the IoT environment.

For the purpose of providing a tailored service recommendation system in the SIoT, authors in \cite{chen2019time}, proposed a time-aware smart object recommendation model to incorporate user preferences over time and the social similarity of smart objects. This model adopts a latent probabilistic approach to learn user preferences and embeds heterogeneous social relationships of smart objects into a shared lower-dimensional space, facilitating estimation of social similarity. The recommendation list is generated using item-based collaborative filtering. Evaluation on real-world datasets showcases the superior recommendation effectiveness of this approach compared to baseline methods. However, the reliance on object usage events for learning user preferences may limit its effectiveness in scenarios where data on object usage is incomplete or unavailable for certain users. Additionally, the model primarily focuses on the "like" property to connect users with services, overlooking essential factors such as user context, preferences, and specific requirements, potentially resulting in less personalized and accurate recommendations.
In another work by Chen \cite{chen2021graph} presented a graph-based framework to address service recommendation challenges within the realm of the Social Internet of Things (SIoT). This framework takes into account both the social relationships among heterogeneous objects in the SIoT and user preferences. To learn user preferences, a latent variable model is utilized, which extracts valuable insights from object usage events. Additionally, a knowledge graph is employed to model the relationships between users, objects, and services. The service recommendation task is treated as a knowledge graph completion problem, with the objective of predicting the "like" property that connects users to services. To validate the proposed model, the researchers have developed a SIoT testbed and conducted experiments, demonstrating the feasibility and effectiveness of the framework in enhancing service recommendation within the SIoT environment. While the graph-based service recommendation framework presented by Chen et al. \cite{chen2021graph} addresses the challenges of service recommendation within the SIoT context, there are certain concerns regarding its scalability and ability to handle the cold start problem. As the number of users, objects, and services increases, the size of the knowledge graph can grow significantly, impacting the efficiency and computational requirements of the recommendation process. Moreover, the framework may struggle to provide accurate recommendations for new users or objects with limited or no usage history, as it relies on user preferences and object usage events.

Zhang et al. \cite{zhang2021mining} designed a recommender system specifically tailored for the SIoT. This system leverages the social dynamics influencing the behavior and interactions of autonomous objects within the SIoT to facilitate optimal pairing of objects and improve recommendation outcomes. However, by heavily relying on social dynamics, other crucial factors such as user preferences and contextual information may be overlooked, leading to less personalized and limited recommendation accuracy. Similarly, Dhelim el al. \cite{vesonet,trust2vec} investigated the integration of social dimension in IoT to enhance the accuracy of trust computing systems.

Given that user ratings or preferences for SIoT services are often implicit, traditional recommendation models such as content-based filtering and collaborative filtering may not be suitable. The diversity of data in the SIoT, spanning various modalities such as images, videos, audio, and text, poses a unique challenge for service recommendation. Each modality can offer valuable insights into user preferences and service characteristics. Consequently, there is a growing need for multi-modal service recommendation approaches that can effectively integrate and leverage information from different modalities. For this, in our recent work \cite{khelloufi2023multi}, we proposed a multimodal service recommendation system for SIoT that considers multiple modalities, such as analyzing images to understand user context or utilizing text data to capture user preferences, we gained a more comprehensive understanding of both device’s and users’ needs and provided more accurate and personalized recommendations. 
The aforementioned works have explored different approaches and techniques to enhance the accuracy, personalization, and efficiency of service recommendations in the SIoT environment. From graph-based frameworks to time-aware models and recommender systems leveraging social dynamics, researchers have made significant strides in addressing the unique challenges posed by the SIoT context. However, there are still certain limitations and areas for improvement in existing approaches. Our proposed work aims to bridge some of these gaps by introducing a contextually aware service recommendation system. By leveraging the contextual representation of device-service pairs and capturing latent feature interactions, we aim to enhance the service recommendation system within the SIoT environment.

\section{System Design and Modeling}
\label{sec:system_design}
\begin{figure*}
\centering
 \noindent \includegraphics*[width=7.00in, height=2.56in]{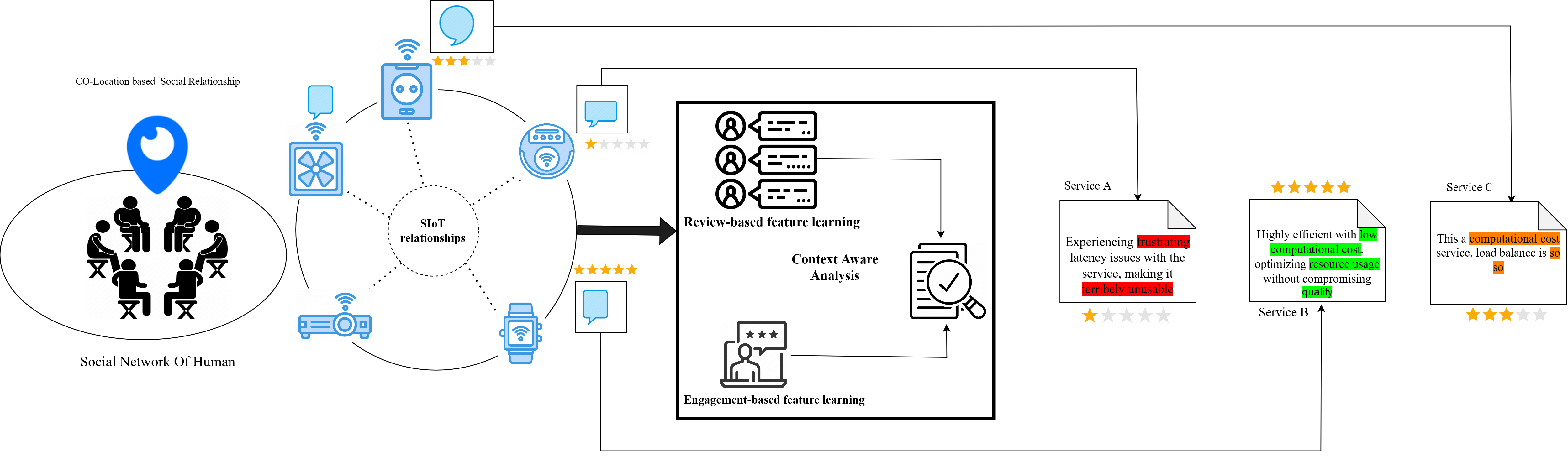}
  \caption{Recommendation Process.}   
\label{image_1}
\end{figure*}

In this section, we introduce our proposed framework, which aims to enhance service recommendation through a context-aware device-service representation learning model. The framework, as depicted in Figure \ref{image_1}, is designed to estimate rating scores for device-service pairs by leveraging two distinct sources of information: service reviews contributed by devices and the device-service interaction matrix. Consequently, the proposed framework comprises two separate feature learning components, namely review-based feature learning and engagement-based feature learning. In the subsequent sections, we provide a comprehensive explanation of the rating prediction framework, followed by a detailed description of the two learning components.

\subsection{Framework Overview}
Traditional recommendation systems commonly employ the dot-product (DP) operation for rating prediction. However, this operation imposes a stringent constraint where the latent dimensions are treated as independent entities. Consequently, each dimension in a latent user vector can only interact with the corresponding dimension in the latent item vector. This independence constraint limits the capability to capture complex user-item rating behaviors that arise from higher-order feature interactions. Considering that our proposed framework is dedicated to SIoT devices that require tailored services, it aims to derive context-aware representations for each device-server pair, it is essential to model these higher-order latent feature interactions to gain a better understanding of rating behaviors.
Therefore, in our approaches, we adopted the Factorization Machine (FM) approach \cite{khelloufi2023multi} to compute the rating score. Specifically, given a learned latent feature vector for a device-service pair, denoted as \(F_{d,s}\), FM calculates the corresponding \(F_{d,s}\) rating score using the following equation:

\begin{equation}\label{eq_1}
\hat{R}_{d,s}(F_{d,s})=b_0+\textbf{a}^{T}_{F_{d,s}}+\frac{1}{2}F^{T}_{d,s}W_{F_{d,s}}
\end{equation}

\begin{equation}\label{eq_2}
W_{j,k}=V^{T}_{j},V_{k}, j\not=k
\end{equation}
Where: \\  
\(b_0\): is the global bias,
\\
\(a\): is the coefficient vector for the latent feature vector \(F_{d,s}\), 
\\
\(W\): is the diagonal weighted matrix \((M_{i,j}=0)\) that represents the interactions of second-order in the system.

The v-dimensional latent vectors $v_{j}$ and $v_{k}$ correspond to dimensions j and k of $F_{d,s}$. 

Equation \ref{eq_1} indicates that both first order $\textbf{a}$ and second-order $M$ feature interactions are employed for predicting ratings.
In traditional recommendation systems, rating patterns encompass various inherent inclinations, referred to as biases. Previous research \cite{koren2009collaborative} has demonstrated that considering user and item biases can effectively account for rating discrepancies and consequently enhance the accuracy of rating prediction. In the context of the SIoT environment, bias referred to the situation in which devices tend to consistently assign higher scores to all services, while certain services are prone to receiving higher ratings from the majority of devices [14]. Building upon these findings, we enhance the rating predictor by incorporating both user and item biases in as follows:

\begin{equation}\label{eq_3}
\hat{R}_{d,s}=\hat{R}_{d,s}(F_{d,s})=b_d+b_i
\end{equation}
In which: $b_d$ and $b_s$ are considered the biases associated with the device and service, respectively. $R_{i,j}$ represents the predicted rating. To optimize the parameters, we utilized the square loss as the objective function. Consequently, we obtained the following equation:
\begin{equation}\label{eq_4}
L_{d,s}=\sum_{(d,s)\in O}^{}(\hat{R}_{d,s}-{R}_{d,s})^2+\lambda_{\Theta}\left\| \Theta \right\|
\end{equation}
Where $O$ represents the set of observed rating pairs for devices and services. ${R}_{d,s}$ denotes the rating score for device $d$ on service $s$, and $\Theta$ represents all the parameters involved. The second term in equation \ref{eq_4} serves as a regularization term, which is employed to prevent the model from overfitting.

\begin{table}
\caption{Preliminaries}
\label{table:demo}
\setlength{\tabcolsep}{3pt}
\begin{tabular}{|p{40pt}|p{180pt}|}
\hline
Symbol& 
Definition \\
\hline
$\hat{R}_{d,s}$& The rating predictor for a device-service pair
\\
$F^{T}_{d,s}$& the latent feature vector
for a device-service pair
\\
$W_{j,k}$& A diagonal weighted matrix
\\
$D_{r}$& Review collection
\\
$M_{d}$& Collection matrix
\\
$L_{d,s}$& Optimized predited rate
\\
$\lambda_{\Theta}$&Regularization term function
\\
$\sigma$&Sigmoid function (non-linear activation function)
\\
$CW_{c}^{j}$&convolution weights associated with filter $f_j$
\\
$ReLU(x)$&Rectified Linear Unit activation function
\\
$g^d_j$ and $g^d_j$& The pair of contextual feature vectors
\\
$R^k_j$& The similarity scores matrix
\\
$tanh$&hyperbolic tangent function
\\
${a}^d$ and ${a}^s$&the selective weight vectors
\\
${D}^w_{h:h+s-1}$&slice of the matrix D obtained by selecting the elements within the sliding window starting at position 
\\
${CW}_{a}^{j}$&the convolution weight vector
\\
$\rho$&A non-linear activation function
\\
${v}_{d}$and ${v}_{s}$& device-service latent vectors
\\
$ F_{collection}$& The feature collection vector
\\
$\beta$&Prediction score parameter
\\
$\oplus$& Element-wise product operation
\\
$e^{collection}_{d,s}$& The feature collection vector
\\

\hline
\end{tabular}
\end{table}
\subsection{Reviews-aggregation process}
\begin{figure}
\centering
 \noindent \includegraphics*[width=2.500in, height=4.56in]{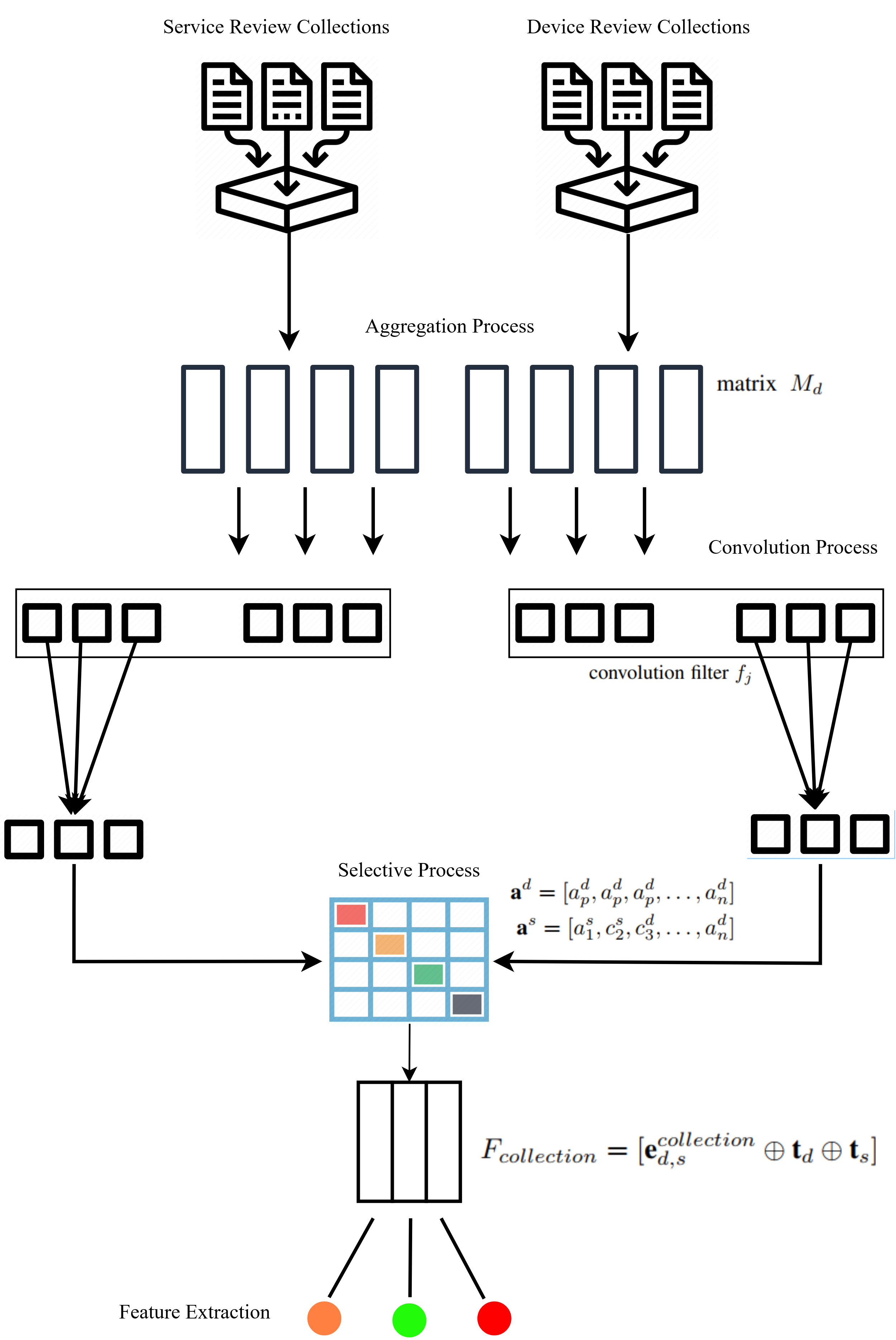}
  \caption{Recommendation Process.}   
\label{image_2}
\end{figure}

In this section, we present the review-based feature learning approach that we employed to enhance the recommendation process. In the context of SIoT environment, we recognize that when a device $d$ reviews a specific service $s$, it reflects the preferences within the ecosystem. Thus, we aggregate all the reviews authored by the same device owners or devices to form a device review collection. Similarly, we merge all the reviews provided by a set of devices for a particular service to construct a service review collection. These two types of review collections are expected to encapsulate distinct semantic information. While a device review collection may encompass personal preferences of users, a service review collection primarily encompasses various aspects that hold significance for all relevant devices. Within our proposed frameworks, the primary objective of review-based feature learning is to deduce a latent feature vector for each device-item pair by jointly considering their respective review collections. Leveraging the convolution operation, which has proven successful in multiple natural language processing and information retrieval tasks such as collection representation learning 
we aim to extract diverse aspects covered within the review collections. Subsequently, a selective layer is employed to emphasize the relevant aspects by taking into account both the device's preferences and the characteristics of the service. Lastly, an abstraction layer is utilized to generate the final latent feature vector for the device-service pair. Figure \ref{image_2} illustrates the network architecture for review-based feature learning.
\subsubsection{Convolution Layer}
In the convolution layer, we process a review collection $D_{r}=\{{x_0....x_{i-1},x_i,x_{i+1}......x_n\}}^{T}$ Initially, a lookup layer maps each word $x_i$ to its respective embedding 
$x_i \in\mathbf{R}^{1 \times T}$.
These embeddings are then concatenated in the order of their appearance within the constructed collection, forming a collection matrix $M_d=\{{x_0....x_{i-1},x_i,x_{i+1}......x_n\}}$ Here, $x_i$ represents the word embedding for the word at the i-th position in collection $D_{r}$. The matrix D maintains the word order, allowing the convolution layer to capture more precise semantic information compared to traditional bag-of-words approaches [35] in traditional recommendation systems. Specifically, we utilize a convolution filter $f_j$ with a sliding window of size s to extract the contextual feature $c^{j}_{h}$ from the local context. By utilizing the convolutional filters with a sliding window, the convolution layer can capture contextual information and extract higher-level features from the input. In the case of review-based feature learning, the convolution layer aims to extract semantic information from the review collection by analyzing the contextual relationships among the words. Specifically:
\begin{equation}\label{eq_5}
c_{h}^{j}=\sigma(CW_{c}^{j}D_{h:h+s-1})
\end{equation}
The function $\sigma$ represents a non-linear activation function. The weight vector $CW_{j}^{c}$ corresponds to the convolution weights associated with filter $f_j$. The notation $D_{p:+s-1}$ represents a slice of the matrix $D$ obtained by selecting the elements within the sliding window starting at position $p$.
By default, we adopt the Rectified Linear Unit (ReLU) as the activation function, defined as $ ReLU(x) = max(0, x)$. To ensure consistent input dimensions, we append $s-1$ zero vectors at the end of the collection matrix $D$. This padding ensures that n contextual features are produced, where $n$ represents the length of the collection $D_r$
To capture various contextual features with service reviews, multiple convolution filters are employed. Each filter utilizes a distinct convolution weight vector to extract contextual features for each word within its local context, spanning s consecutive words. Specifically, in this study, we employ two different sets of convolution weight vectors $CW^{\ast }_{c}$, to process the user review collection and the item review collection separately.

\subsubsection {Selective Layer}
Building on our previous discussion, we acknowledge that a device/user review collection which we referred as review collections has the potential to encompass personalized and diverse preferences for different services. Similarly, a service review collection may focus on distinct aspects that different users find important. Consequently, not all the information present in the review collections is necessarily relevant for predicting the rating score of a particular device-service pair. To effectively capture the useful information, we incorporate a selective layer that operates on the review collections corresponding to the specific device-service pair.
After applying the convolution operation to a review collection, we obtain a contextual feature vector $c_p$, which represents the features extracted for the word located at the p-th position within the service review collection.
After applying the convolution operation to a review collection, we obtain a contextual feature vector $c_p$, which represents the features extracted for the word located at the p-th position within the service review collection. Therefore, we can establish two matrices D and S to represent the devices and services review collections. where $c^d_j$ and $c^s_k$ correspond to the contextual feature vectors for the j-th word and the k-th word in the device and service review collections, respectively.
  \begin{equation}\label{eq_6}
\left\{
\begin{matrix} 
& c_p = [c^1_p,c^2_p,c^3_p ....... , c^{f-1}_p,c^f_p] \\
& \Downarrow  \\
& D = [c^d_1,c^d_2,c^d_3 ....... , c^d_{n-1},c^d_n] \\ 
& S = [c^s_1,c^s_2,c^s_3 ....... , c^s_{n-1},c^s_n]
\end{matrix}
\right.
\end{equation}
In order to determine the significance of each contextual feature vector for both D and S we introduce a selective matrix denoted as $A \in\mathbf{R}^{f \times f}$, Specifically, we project matrices D and S into a shared latent space and compute the pairwise relevance between each pair of contextual feature vectors, $c^d_j$ and $c^s_k$, using the following approach:
\begin{equation}\label{eq_7}
    R^k_j = \tanh(c^d_jAc^s_k)
\end{equation}

According to equation \ref{eq_7}, each row  $R^j_{ \ast}$ in a column $R^{\ast}_k$ represents the similarity scores between the vectors $c^d_j$ and $c^s_k$ in matrices U and S respectively. To calculate the similarity scores, a mean-pooling operation is performed on each row/column of the matrix R, which is mathematically expressed as:
\begin{equation}\label{eq_8}
\begin{matrix}
& g^d_j=\textbf{mean}(R^j_1, \dots, R^j_m) \\ 
& \\
& g^s_j=\textbf{mean}(R^1_k\dots R^n_k) 
\end{matrix}
\end{equation}

\begin{equation}\label{eq_9}
\begin{split}
l^d_j= \frac{\exp(g^d_j)}{\sum_{h}^{n}\exp(g^d_h)} \\
\\ l^s_j=\frac{\exp(g^s_k)}{\sum_{h}^{m}\exp(g^s_h)}
\end{split}
\end{equation}

\begin{equation}\label{eq_10}
\begin{split}
\textbf{a}^d = [a^d_p,a^d_p,a^d_p, \dots , a^d_n] \\ 
\textbf{a}^s= [a^s_1,c^s_2,c^d_3, \dots ,a^d_n] 
\end{split}
\end{equation} 
By utilizing the mean similarity computed in equations \ref{eq_8}, we can determine the significance of each contextual feature vector in D and S. This is expressed through the selective weights $l^d_j$ and $l^s_k$, as shown in equation \ref{eq_9}. Ultimately, the selective layer yields the selective weight vectors $\textbf{a}^d$ and $\textbf{a}^s$, as demonstrated in equation \ref{eq_10}. These weight vectors represent the learned distribution indicating the level of importance assigned to the words in the device review collection and service review collection, respectively.

\subsubsection{Feature Extraction Layer}
In this subsection, we describe the process of extracting higher-level semantic features from device and service review collection using an abstraction layer. First, selective vectors represented as $\textbf{a}^d$ and $\textbf{a}^s$ are calculated based on the device and service collections. These attentive weights indicate the relevance of each contextual feature vector to the device-service pair. A higher weight signifies greater relevance (equation \ref{eq_11}).
While it is possible to represent the device and service by simply summing up the weighted contextual feature vectors, such a simplistic approach may introduce excessive noise due to the inclusion of irrelevant aspects from both review collections. To overcome this limitation, we adopt a more sophisticated approach by employing further neural transformations (a mean-pooling CNN network) based on the weighted contextual vectors $D^w$ and $S^w$.
One advantage of applying a mean-pooling CNN network over max-pooling on $D^w$ and $S^w$ is that it enables the extraction of latent features based on a larger context while taking into account the relevance weights as depicted in equation \ref{eq_12}.  Moreover, the mean-pooling strategy is selected over max-pooling due to its ability to capture the diversity of devices/user opinions across different aspects of service. For instance, when selecting a streaming service by a smart TV, consider not only the available content library but also the user interface, streaming quality, and subscription plans to enhance the user’s entertainment experience. \\ Equation \ref{eq_13} describes the process. Where $\sigma$ is sigmoid function applied along with the convolution weight vector $\textbf{CW}_{a}^{j}$ and device item collection matrix $\textbf{D}^w_{h:h+s-1}$.
After the mean-pooling step, a shared MLP layer is stacked upon the mean-pooled vectors. As depicted in \ref{eq_13} The vectors $H^d$ and $H^s$ are transformed using the transformation matrix $\textbf{CW}_1$ and bias vector $\textbf{b}_1$ through the sigmoid function $\sigma$.The transformed vectors $t_d$ and $t_s$ represent the devices and services specific transformed features respectively. 

The context-aware latent feature vector for the device-service pair is then formed by combining the review embedding, device-specific features $t_d$, and service-specific features $t_d$. This is illustrated in \ref{eq_14}. The element-wise product enhances the interactions between the latent features. The resulting vector i.e. the feature collection vector, denoted as $\textbf{Fcollection}$, captures both the individual characteristics and their interactions, providing a comprehensive representation of the device-item pair. 
Overall, this abstraction layer takes the weighted contextual vectors, applies a mean-pooling CNN network to extract higher-level semantic features, and then employs a shared MLP layer for further feature extraction.

\begin{equation}\label{eq_11}
\begin{split}
\textbf{D}^w =\textbf{a}^d\textbf{D} \\ 
\textbf{S}^w= \textbf{a}^s\textbf{S}
\end{split}
\end{equation}
 
\begin{equation}\label{eq_12}
\left\{
\begin{matrix}
 H^j_h=\sigma(\textbf{CW}_{a}^{j}\textbf{D}^w_{h:h+s-1}) \\ \\
H_j=\textbf{mean}(H^j_1,\dots,H^j_n) \\ \\
\textbf{H}^d=[H_1,\dots,H_f] \\
\end{matrix}
\right.
\end{equation}

\begin{equation}\label{eq_13}
\begin{split}
t_d=\sigma(\textbf{CW}_{1}H^{d}+b_1)\\
t_s=\sigma(\textbf{CW}_{1}H^{s}+b_1) \\
\end{split}
\end{equation}

\begin{equation}\label{eq_14}
\begin{split}
 F_{collection}=[\textbf{e}^{collection}_{d,s}\oplus \textbf{t}_d\oplus \textbf{t}_s]
\end{split}
\end{equation}

\subsection{Engagement Feature learning Process}
In order to capture the device’s rating behaviors more comprehensively, we incorporate engagement-based feature learning. While textual reviews provide valuable information, the previously obtained review-based latent features alone may not fully represent the device’s rating preferences. To overcome this, we introduce a separate set of latent vectors for devices and services. The devices and service identities are one-hot encoded and mapped onto their respective latent vectors $\textbf{v}_d$ and $\textbf{v}_s$ using matrices $\textbf{D}$ and $ \textbf{Q}$
Since no contextual information like textual reviews is available, an element-wise product operation is used to extract engagement-based features. equations \ref{eq_15} and \ref{eq_16} depicted the process of engagement-based Feature extraction. 
\begin{equation}\label{eq_15}
\begin{split}
 \textbf{v}_d = \textbf{D}_{x_d} \\
\textbf{v}_s = \textbf{S}_{x_d}
\end{split}
\end{equation}
\begin{equation}\label{eq_16}
\begin{split}
\rho = \textbf{v}_d \circ    \textbf{v}_{s} \\
F_{engagement} = [ \rho \oplus \textbf{v}_d \oplus \textbf{v}_{s}]
\end{split}
\end{equation}
\subsection{Feature learning combination}
The combination process aims to integrate the context-aware latent features extracted from the two aforementioned processes. These include the review-based features $F_{collection}$ and the engagement-based features $F_{engagement}$. The objective is to leverage the strengths of both components and enhance prediction performance.
A simple approach is to linearly interpolate the estimated rating score $\hat{R}_{d,s}$ by combining the two components. This is achieved using a parameter $\beta$, which serves as a tradeoff between the review-based and engagement-based components. However, a static $\beta$ value may not be suitable since users/devices may assign higher ratings based on specified preferred features while neglecting others.
To address this, we introduce a dynamic weighting scheme. We calculate $\beta$ by comparing the prediction scores of the review-based component $\hat{R}_{d,s}(F_{collection})$  and the engagement-based component $\hat{R}_{d,s}(F_{engagement})$. The parameter $\beta$ is then determined based on the relative rating predictions, ensuring that the component with the higher score is given more importance in the final rating prediction.
In summary, the engagement-based feature learning process focuses on capturing features from device-service interactions, while the combination step allows merging these features with the review-based features. The dynamic weighting scheme enables adaptive weighting based on the predictive power of each component, resulting in improved rating prediction accuracy (equation \ref{eq_17}).

\begin{equation}\label{eq_17}
    \begin{aligned}
\hat{R}_{d,s} &= \beta\hat{R}_{d,s}(F_{collection}) \\
&\quad+(1-\beta)\hat{R}_{d,s}(F_{engagement}) \\
&\quad+b_d+b_s \\
\beta &= \frac{\beta\hat{R}_{d,s}(F_{collection})}{\beta\hat{R}_{d,s}(F_{collection})+(\hat{R}_{d,s}F_{engagement})}
\end{aligned}
\end{equation}

\begin{table}
\small
\centering
\caption{Summary of the Amazon Review Dataset}
\label{tab:datasetsummary}
\begin{tabular}{@{}lcccc@{}}
\toprule
Dataset & Users & Items & Ratings & Density \\
\midrule
Appliances & 1,565 & 1052 & 11,342 & 79.8\% \\
Cellphones & 6,019 & 3,281 & 66,239 & 44.8\% \\
Electronics & 7,529 & 3,657 & 39,256 & 32.7\% \\
\bottomrule
\end{tabular}
\end{table}
\section{Experimental Evaluations}
\label{sec:evaluation}

\subsection{Dataset Description}
Due to the lack of publicly available SIoT datasets, we utilized the Amazon review dataset \cite{pennington2014proc} for the performance evaluation of the proposed service recommendation. This dataset, summarized in Table \ref{tab:datasetsummary}, provides a comprehensive collection of reviews from various product categories. Although it is not specific to the SIoT domain, it offers a valuable resource for evaluating the effectiveness of the recommendation approaches. The dataset encompasses a wide range of user reviews and corresponding ratings, which allow us to assess the performance and compare the proposed methods against existing approaches.

\subsection{Evaluation Methodology}
In this section, we present a comprehensive evaluation methodology to assess the performance and effectiveness of the proposed recommendation approach. We outline the metrics used as baselines for comparing and evaluating the proposed framework. The evaluation focuses on various metrics that measure the quality of recommendations and their ability to address the unique challenges posed by the SIoT environment. The baseline metrics considered include recall, precision, Normalized Discounted Cumulative Gain (NDCG), Root Mean Squared Error (RMSE), and Mean Absolute Error (MAE). These metrics provide valuable insights into the accuracy, relevance, and ranking of the recommended items. By considering these metrics, we provide a comprehensive performance evaluation and comparison of our proposed approaches with existing recommendation methods such as Matrix Factorization (MF), Graph-Based Service Recommendation (GBSR), Smart Object Recommendation Framework (BLA), Object Recommendation with Text-Topic Information (ORTJ), and Multi-Modal Recommendation System (MMRS).

\subsubsection{Baseline}
The baseline approaches discussed in this section provide reference models for evaluating the proposed recommendation methods. They encompass Matrix Factorization (MF), Graph-Based Service Recommendation (GBSR), Smart Object Recommendation Framework (BLA), Object Recommendation with Text-Topic Information (ORTJ), and Multi-Modal Recommendation System (MMRS).

\begin{itemize}

\item MF: It is a collaborative filtering method used to enhance recommendation accuracy by reducing data sparsity in the user-item matrix\cite{koren2009matrix}. It creates user and item matrices, where each row/column represents a vector for the corresponding user/item. The predictive score of user i for item j is calculated by multiplying the corresponding vectors in the two matrices. The matrices are adjusted during training to minimize the least squared error between the actual and predicted values.
\item GBSR: It is a framework proposed to address the service recommendation problem in SIoT\cite{chen2021graph}. It models the SIoT service recommendation as a knowledge graph completion problem. The framework leverages the user's preferences and object usage events, including rich spatial-temporal information, to uncover the user's preferences. Specifically, it models the hidden factors of the user's object usage and constructs a knowledge graph based on the service usage value, which is reflected by the service usage frequency. 

\item BLA: It is a framework designed to recommend smart objects based on user requirements in the SIoT \cite{zhang2021mining}. This model utilizes Bi-LSTM and BERT to generate vectors for matching and recommendations. It employs self and global attention mechanisms to dynamically adjust the weights of vectors for improved performance. Additionally, the model incorporates Thing-thing relationship data to better utilize user requirements in generating reasonable representations of smart object attributes and characteristics. The authors evaluate the model using an extended original MovieLens dataset in SIoT scenarios.

\item ORTJ: It is a proposed approach for recommending smart objects in the IoT \cite{zhang2022smart}. It considers both the attributes and text-topic information of the objects. To enhance recommendation accuracy, a "thing-thing" relationship is introduced as an attribute for smart objects. The ORTJ model, based on maximum a posteriori estimation, is developed, and experiments are conducted to compare its performance with other models.
\item MMRS: It is one of our previous works in the service recommendation for SIoT. MMRs is a multi-modal service recommendation system that considers the diversity of data generated in the SIoT environment \cite{khelloufi2023multi}. The proposed system analyses the multimodal features such as item-item relationships to provide tailored service recommendations in SIoT environment. It provided an adaptive service recommendation system that can learn from item-item structure and improve the accuracy of future recommendations.
\end{itemize}
\subsubsection{Metrics}
In order to conduct a comprehensive evaluation of the proposed framework, we employed a set of widely used and effective evaluation metrics to measure its effectiveness. These metrics provide valuable insights into the performance of the framework and facilitate a thorough assessment of its recommendation quality. The evaluation metrics we considered include precision, recall, normalized discounted cumulative gain (NDCG), mean absolute error (MAE), and root mean squared error (RMSE).
\begin{itemize}

\item Precision is a metric that measures the accuracy of the recommended items. It quantifies the proportion of relevant items among the recommended ones. Recall, on the other hand, evaluates the coverage of the recommendation system by measuring the proportion of relevant items that have been successfully recommended.

\item NDCG is a metric that takes into account both the relevance and ranking position of recommended items. It considers the graded relevance of the items and assigns higher importance to those items that are ranked higher in the recommendation list.

\item MAE measures the average magnitude of the errors between the predicted ratings and the actual ratings. It provides a straightforward evaluation of the absolute accuracy of the recommendations.

\item  RMSE, similar to MAE, assesses the accuracy of the recommendations by measuring the square root of the average squared differences between the predicted ratings and the actual ratings. It penalizes larger errors more heavily than MAE.
 \begin{equation}\label{equation _18}
\begin{aligned}
\textbf{Recall@k} &= \frac{\text{Correctly Recommended Services}}{\text{Total Relevant}} \\ &= \frac{1}{n}\sum_{i=1}^{n}\frac{TP_{i}}{TP_{i}+FN_{i}} \\
\textbf{Precision@k} &= \frac{\text{Correctly Recommended Services}}{\text{Total Recommended}} \\ &= \frac{1}{n}\sum_{i=1}^{n}\frac{TP_{i}}{TP_{i}+FP_{i}} \\
\textbf{MAE} &= \frac{1}{n}\sum_{i=1}^{n}\ \left| \textbf{Pr}_i-\textbf{R}i \right| \\
\textbf{RMSE}&=\sqrt[2]{\frac{\sum_{1}^{n}(\left| \textbf{Pr}_i-\textbf{R}_i  \right|)}{n}}
\end{aligned}
\end{equation}
 
\item NDCG
The Normalized Discounted Cumulative Gain (NDCG) is a widely used evaluation metric in recommendation systems that measures the quality of the recommended items, taking into account both their relevance and their ranking position in the recommendation list. The equation for NDCG is as follows:

\[ \text{NDCG@k} = \frac{\text{DCG@k}}{\text{IDCG@k}} \]
Where:
- NDCG@k is the NDCG value at position \( k \).
- DCG@k is the Discounted Cumulative Gain at position \( k \).
- IDCG@k is the Ideal Discounted Cumulative Gain at position \( k \).
The DCG@k is calculated by summing up the graded relevance of the recommended items at each position, discounted logarithmically based on their position:\\
\[ \text{DCG@k} = \sum_{i=1}^{k} \frac{\text{rel}_i}{\log_2(i+1)} \]
Where \( \text{rel}_i \) represents the graded relevance of the item at position \( i \).
The IDCG@k represents the ideal DCG value at position \( k \), which is obtained by sorting the relevance values in descending order and calculating the DCG@k based on this ideal ranking.
Once DCG@k and IDCG@k are calculated, the NDCG@k is obtained by dividing DCG@k by IDCG@k.
\[ \text{NDCG@k} = \frac{\text{DCG@k}}{\text{IDCG@k}} \]

The NDCG value ranges between 0 and 1, where 1 represents the highest quality of recommendations, indicating that the recommended items are highly relevant and well-ranked in the list.
\end{itemize}

\subsubsection{Evaluation Details}
In this section, we present the evaluation details of our context-aware service recommendation system for the Social Internet of Things. We aim to address three specific research questions that pertain to the performance and effectiveness of our system. These questions are as follows:
\begin{itemize}
\item \underline{\textbf{RQ1}}: How does combining review-based feature learning and engagement-based feature learning enhance the performance of the service recommendation system in the social Internet of Things?
\item \underline{\textbf{RQ2}}: What is the impact of different hyperparameter settings on the accuracy and effectiveness of the service recommendation system in SIoT?
\item \underline{\textbf{RQ3}}: How effectively does the selective layer in the review-based feature learning component identify relevant semantic information from devices-service pairs ?
\end{itemize}

 \begin{figure*}
\centering
 \noindent \includegraphics*[width=7.00in, height=2.56in]{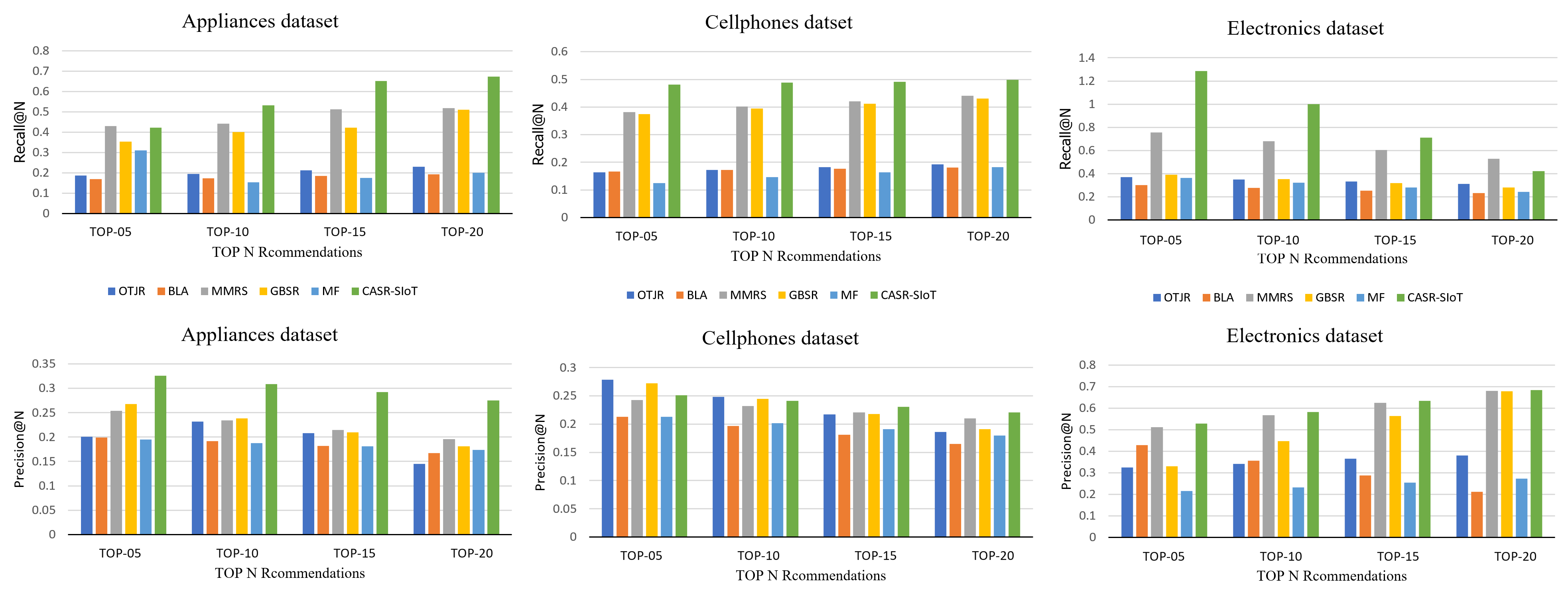}
  \caption{Results of the precision and recall with different baseline on 3 different datasets}   
\label{image_3}
\end{figure*}
\subsubsection{Performance Comparison (\textbf{RQ1})}
To address our first research question, we conducted experiments using the aforementioned three categories of datasets. In order to establish a comparison, we evaluated the performance of the proposed framework against various baseline methods. In the following subsection, we present a comprehensive performance comparison in terms of recall and precision for the Appliances, Cellphones, and Electronics categories.

As shown in \ref{image_3}, the performance comparison clearly indicates that the proposed framework consistently outperforms the baseline methods across all three datasets (Appliances, Cellphones, and Electronics) in terms of both recall and precision. Moving forward, we delve into a detailed analysis of the recall and precision metrics for the Appliances, Cellphones, and Electronics datasets, focusing specifically on the performance comparison between the proposed framework and the baseline methods for the service recommendation system in the social domain.

Our proposed framework, CASR-SIoT, demonstrates superior performance across all aspects of the service recommendation system in the SIoT domain. By combining review-based feature learning and engagement-based feature learning, which enables context-aware analysis, CASR-SIoT consistently achieves higher recall and precision values. This indicates its effectiveness in retrieving relevant and accurate services for devices. Furthermore, the results highlight that the MMRS method, incorporating multi-modal features, performs competitively in terms of recall and precision. However, the MF (Matrix Factorization) approach, relying solely on matrix analysis, falls short in comparison to the proposed framework.

In summary, our findings consistently demonstrate that the proposed framework outperforms the baseline methods in terms of both recall and precision across all three categories. This compelling evidence indicates that the integration of review-based feature learning and engagement-based feature learning significantly enhances the performance of the service recommendation system in the social domain. By effectively retrieving more relevant services and providing more accurate recommendations, the proposed framework surpasses the capabilities of the baseline methods.

\begin{table*}
  \centering
  \caption{Performance comparison of the proposed recommendation system compared to baseline in terms of Recall@15, Precision@15, and RMSE@5. The improvement line represents the percentage of relative enhancements compared to the best baseline}
  \label{tab:results-01}
  \begin{tabular}{cccccccccc}
    \hline
    \multirow{2}{*}{Model} & \multicolumn{3}{c}{Appliances} & \multicolumn{3}{c}{Cell Phones} & \multicolumn{3}{c}{Electronics} \\ \cline{2-10}
                                     & \textbf{R@15} & \textbf{P@15} & \textbf{RMSE@5} & \textbf{R@15} & \textbf{P@15} & \textbf{RMSE@5} & \textbf{R@15} & \textbf{P@15} & \textbf{RMSE@5} \\ \hline
    MF                               & 0.175         & 0.181         & 0.981           & 0.163         & 0.191         & 1.502           & 0.253         & 0.282         & 1.432           \\ \hline
    GBSR                             & 0.423         & 0.098         & 1.982           & 0.412         & 0.218         & 2.121           & 0.563         & 0.318         & 1.932           \\ \hline
    BLA                              & 0.185         & 0.182         & 0.891           & 0.176         & 0.181         & 1.253           & 0.284         & 0.254         & 1.142           \\ \hline
    ORTJ                             & 0.212         & 0.208         & 0.980           & 0.182         & 0.217         & 1.258           & 0.962         & 0.331         & 1.001           \\ \hline
    MMRS                             & 0.513         & 0.215         & 0.781           & 0.421         & 0.221         & 0.889           & 0.621         & 0.303         & 0.745           \\ \hline
    CASR-SIOT                        & 0.652         & 0.292         & 0.632           & 0.492         & 0.231         & 0.874           & 0.631         & 0.411         & 0.695           \\ \hline
    \rowcolor[gray]{0.9} Improvement & \textbf{14\%} & \textbf{8\%}  & \textbf{15\%}   & \textbf{7\%}  & \textbf{1\%}  & \textbf{2\%}    & \textbf{1\%}  & \textbf{11\%} & \textbf{5\%}    \\ \hline
  \end{tabular}
\end{table*}

\subsubsection{Different models and hyperparameters comparison (\textbf{RQ2})}
\begin{figure}
\centering
 \noindent \includegraphics*[width=2.50in, height=2.5in]{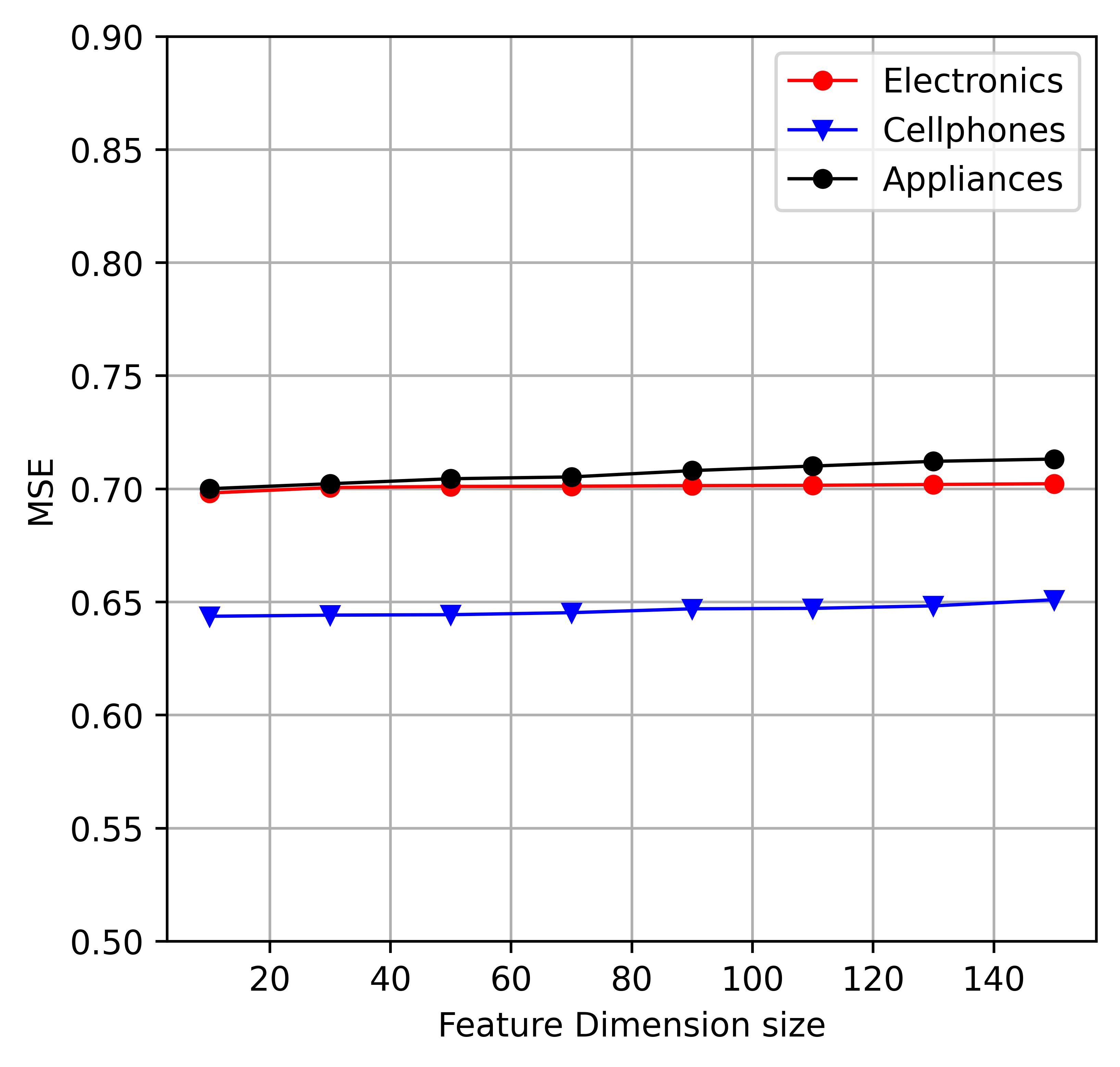}
  \caption{MSE values variations according to the Feature dimension size with different datasets}   
\label{image_4}
\end{figure}

In this section, we aim to address the second research question by conducting a comparison of various models with different feature dimension size parameters. The objective is to validate the performance of the proposed frameworks under different settings. We investigate the impact of altering the feature dimension size and explore the use of Linear Regression as an alternative to the factorization machine model. By separating the models, we can thoroughly evaluate their respective performances and make meaningful comparisons.
\paragraph {Feature dimension size}
In order to evaluate the performance of the proposed framework, we calculated the mean absolute error (MAE) and examined its performance across different dataset categories. Figure \ref{image_4} clearly demonstrates the consistent outperformance of the proposed framework compared to the baseline methods for all three categories (electronics, cellphones, appliances). Notably, even with a threshold value of 10, the proposed framework exhibited strong performance across all categories. It is worth noting, however, that higher values of the feature dimension parameter (f) showed improved performance at the expense of increased computational costs, given the utilization of the Factorization Machine method. To balance performance and computational efficiency, we decided to fix the feature dimension at $f=20$ for the subsequent analysis and evaluation of the proposed framework's performance.

\paragraph{Factorization Machine and different Models Investigation}
The impact of using the Factorization Machine (FM) method compared to the commonly used Linear Regression, as well as the investigation of the effectiveness of models separately, were studied to evaluate the performance of the recommendation system in SIoT environment in the context of review-based and engagement-based learning. The study aimed to determine which method yields better results in terms of accuracy and the obtained results are summarized in Table \ref{tab:rating-prediction}. 
The table provides a comparison of different methods used for prediction in three categories: Electronics, Cellphones, and Appliances. Each method is evaluated based on the Mean Absolute Error (MAE) metric, which measures the average absolute difference between the predicted values and the actual values. Among the models, the rating-based alone demonstrates the highest MAE values in the all-dataset categories, indicating its lower accurately predicting ratings for these types of products when it is solely relies on ratings. The review-based method performs better than rating based approaches but shows slightly higher MAE values for the dataset appliances. Which proves that review provided for this category of dataset are not well relevant.  On the other hand, the LR-based method and FM-based (Factorization Machine-based) method have lower MAE values across all three categories. However, it is important to note that the FM-based method shows better performance than the LR-based method in terms of MAE values. To quantify the difference in performance between FM and LR, the percentage difference of MAE values between the FM-based method and the other methods shows that the FM-based method has a 12.50\%, 19.46\%, and 13.97\% lower MAE values for Electronics, Cellphones, and Appliances respectively. Therefore, which the accuracy results across all dataset categories, showcases the superiority of FM over Linear Regression. The FM method consistently achieves higher accuracy values, highlighting its capability to generalize well to unseen data and handle feature engagement that are not explicitly specified. Moreover, FM demonstrates its effectiveness in handling sparse and high-dimensional data, which is particularly important in the context of these datasets.
The results clearly demonstrate that employing the FM method outperforms the traditional Linear Regression approach across all three categories of datasets. FM, known for its flexibility and ability to capture complex feature interactions, proved to be highly effective in this study. By decomposing feature interactions into latent factors through factorization, FM can capture both pairwise and higher-order interactions, leading to improved accuracy.
In contrast, Linear Regression assumes a linear relationship between the features and the target variable, disregarding complex interactions or non-linear patterns. Although Linear Regression is computationally efficient and suitable for cases where the relationship between features and the target variable can be approximated by a linear function, it falls short in capturing the intricate feature interactions present in the datasets considered in this study.
 
\begin{table}[htbp]
\small
\centering
\caption{Comparison of Methods for Rating Prediction}
\label{tab:rating-prediction}
\begin{tabular}{@{}lccc@{}}
\toprule
\textbf{Method} & \textbf{Electronics} & \textbf{Cellphones} & \textbf{Appliances} \\ 
\midrule

Review-based & 0.825 & 0.923 & 1.020 \\ 
Engagement-based & 0.831 & 0.812 & 0.874 \\ 
LR-based & 0.801 & 0.792 & 0.846 \\ 
FM-based & \textbf{0.712} & \textbf{0.663} & \textbf{0.7423} \\ 
\rowcolor[gray]{0.9} Percentage Diff. & 12.50\% & 19.46\% & 13.97\% \\ 
\bottomrule
\end{tabular}
\end{table}

\subsubsection{Different models and hyperparameters comparison (\textbf{RQ2})}
 In this section, we focus on the impact of the selective layer within the review-based feature learning component, specifically for device-service pairs, to identify relevant semantic information effectively. 
In the proposed recommendation system, the review-based learning component extracts latent features from user reviews and service documents to capture semantic interactions between devices and services. This process contributes to the creation of device-service representations.
To test the effectiveness of the selective layer, we conducted a removal experiment in which we eliminate the selective layer and conduct the review-based approach without it. The results of the experiment are shown in Table \ref{tab:rating-prediction}.
As It is clearly shown that the removal of these element-wise product-based latent features from the review and rating learning components results in performance degradation, signifying their relevance in identifying relevant semantic information.

\section{Conclusion}
The emergence of the Social Internet of Things (SIoT) has revolutionized the way interconnected smart devices share data and services, leading to the development of personalized service recommendations. Despite these advancements, existing research in this field has often overlooked crucial aspects that could significantly enhance the accuracy and relevance of recommendations within the SIoT context. Specifically, while some techniques have considered social relationships between devices, they have failed to adequately incorporate contextual presentation of service reviews.

This research article aims to address these limitations by introducing a context-aware service recommendation system tailored for the SIoT environment. Our investigation revolves around three primary research objectives: firstly, assessing the impact of combining review-based feature learning and engagement-based feature learning on the performance of the service recommendation system; secondly, examining the influence of different hyperparameter settings on the accuracy and effectiveness of the recommendation system; and lastly, evaluating the effectiveness of the selective layer in review-based feature learning in identifying relevant semantic information from device-service pairs.

To achieve these objectives, we conducted experiments using three categories of datasets and conducted a comprehensive performance comparison of the proposed framework, CASR-SIoT, against various baseline methods. The results consistently demonstrated that CASR-SIoT outperformed the baseline methods across all three datasets, showcasing superior recall and precision values for Appliances, Cellphones, and Electronics categories. The integration of review-based and engagement-based feature learning facilitated context-aware analysis, resulting in more precise and relevant service recommendations.

Our research findings hold significant implications for the domain of SIoT service recommendation systems. Existing models have encountered challenges in adapting to the dynamic nature of the SIoT environment, often relying on limited data sources and overlooking diverse content and modalities. The success of CASR-SIoT in enhancing recommendation accuracy underscores the potential for incorporating context-awareness in future SIoT service recommendation frameworks, offering personalized and contextually relevant service suggestions.

Despite the notable performance improvements achieved by the proposed framework, we acknowledge certain limitations and identify potential future directions. The current study centered on review-based and engagement-based feature learning; future research could explore the integration of additional contextual factors, such as user preferences, location, and time, to further refine the recommendation system. Moreover, incorporating multi-modal data and evolving user interactions could enhance the system's ability to deliver real-time, personalized, and accurate recommendations.

From the results in the previous section, it is shown that this research significantly contributes to the ongoing efforts to optimize SIoT service recommendation systems. By effectively leveraging contextual data and capturing latent feature interactions, CASR-SIoT showcases the potential to provide more accurate, tailored, and relevant service recommendations within the SIoT environment. As the SIoT ecosystem continues to evolve, our study sheds light on the importance of context-awareness in enhancing the user experience and extending the capabilities of the Social Internet of Things.

\label{sec:Conclusion}

\bibliographystyle{IEEEtran}
\bibliography{References}

\vskip 0pt plus -1fil
\begin{IEEEbiography}[{\includegraphics[width=1in,height=1.25in,clip,keepaspectratio]{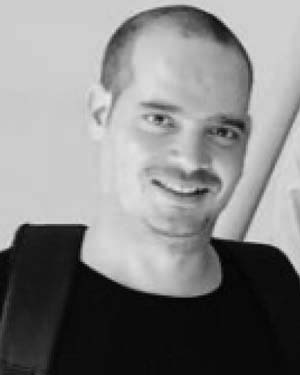}}]{Amar Khelloufi}
Received the B.S. degree (Hons.) in computer science from the Faculty of Sciences and Technology, Ziane Achour University of Djelfa, Djelfa, Algeria, in 2012, and the M.S. degree in distributed information systems from the Faculty of Sciences, University of Boumerdès, Boumerdès, Algeria, in 2014. He is currently pursuing the Ph.D. degree with the School of Computer and Communication Engineering, University of Science and Technology Beijing, Beijing, China. His current research focuses on Internet of Things, blockchain applications, edge computing, and distributed systems.
\end{IEEEbiography}

\vskip 0pt plus -1fil
\begin{IEEEbiography}[{\includegraphics[width=1in,height=1.25in,clip,keepaspectratio]{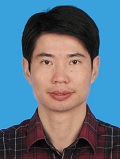}}]{Huansheng Ning}
Received his B.S. degree from Anhui University in 1996 and his Ph.D. degree from Beihang University in 2001. Now, he is a professor and vice dean of the School of Computer and Communication Engineering, University of Science and Technology Beijing, China. His current research focuses on the Internet of Things and general cyberspace. He has presided many research projects including Natural Science Foundation of China, National High Technology Research and Development Program of China (863 Project). He has published more than 200 journal/conference papers, and authored 5 books. He serves as an associate editor of IEEE Systems Journal (2013-Now), IEEE Internet of Things Journal (2014-2018), and as steering committee member of IEEE Internet of Things Journal (2016-Now).
\end{IEEEbiography}

\vskip 0pt plus -1fil
\begin{IEEEbiography}[{\includegraphics[width=1in,height=1.25in,clip,keepaspectratio]{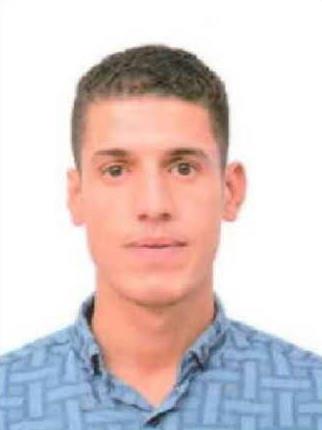}}]{Abdenacer Naouri}
He is currently a Ph.D. candidate at the University of Science and Technology Beijing China, Beijing, China. He 
received his B.S. degree in computer science from the University of Djelfa Algeria, in 2011, and the  M.Sc. degree in networking and distributed systems from the University of Laghouat Algeria, Laghouat, Algeria, in 2016. His current research interests include Cloud computing, Smart communication, machine learning, Internet of vehicles  and Internet of Things. \end{IEEEbiography}

\vskip 0pt plus -1fil
\begin{IEEEbiography}[{\includegraphics[width=1in,height=1.25in,clip,keepaspectratio]{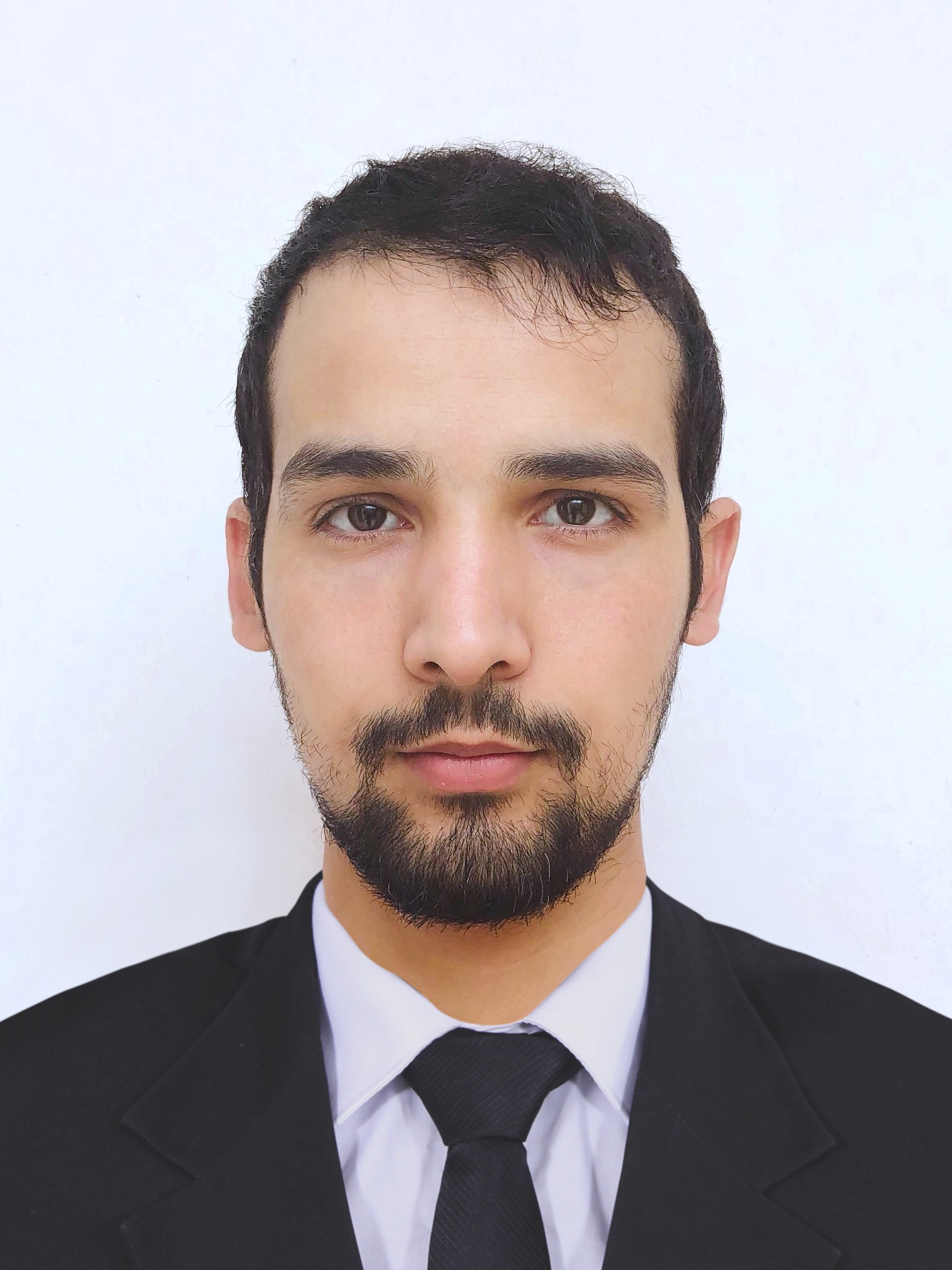}}]{Abdelkarim Ben Sada}
Received his BSc in Computer Science in 2014 from the University of Djelfa Algeria, and his MSc degree in 2016 majoring in Networking and Distributed Systems from the University of Laghouat Algeria. He is currently pursuing his PhD degree at the University of Science and Technology Beijing China. His research interests include Computer Vision, Machine Learning and Internet of Things.
\end{IEEEbiography}

\vskip 0pt plus -1fil
\begin{IEEEbiography}[{\includegraphics[width=1in,height=1.25in,clip,keepaspectratio]{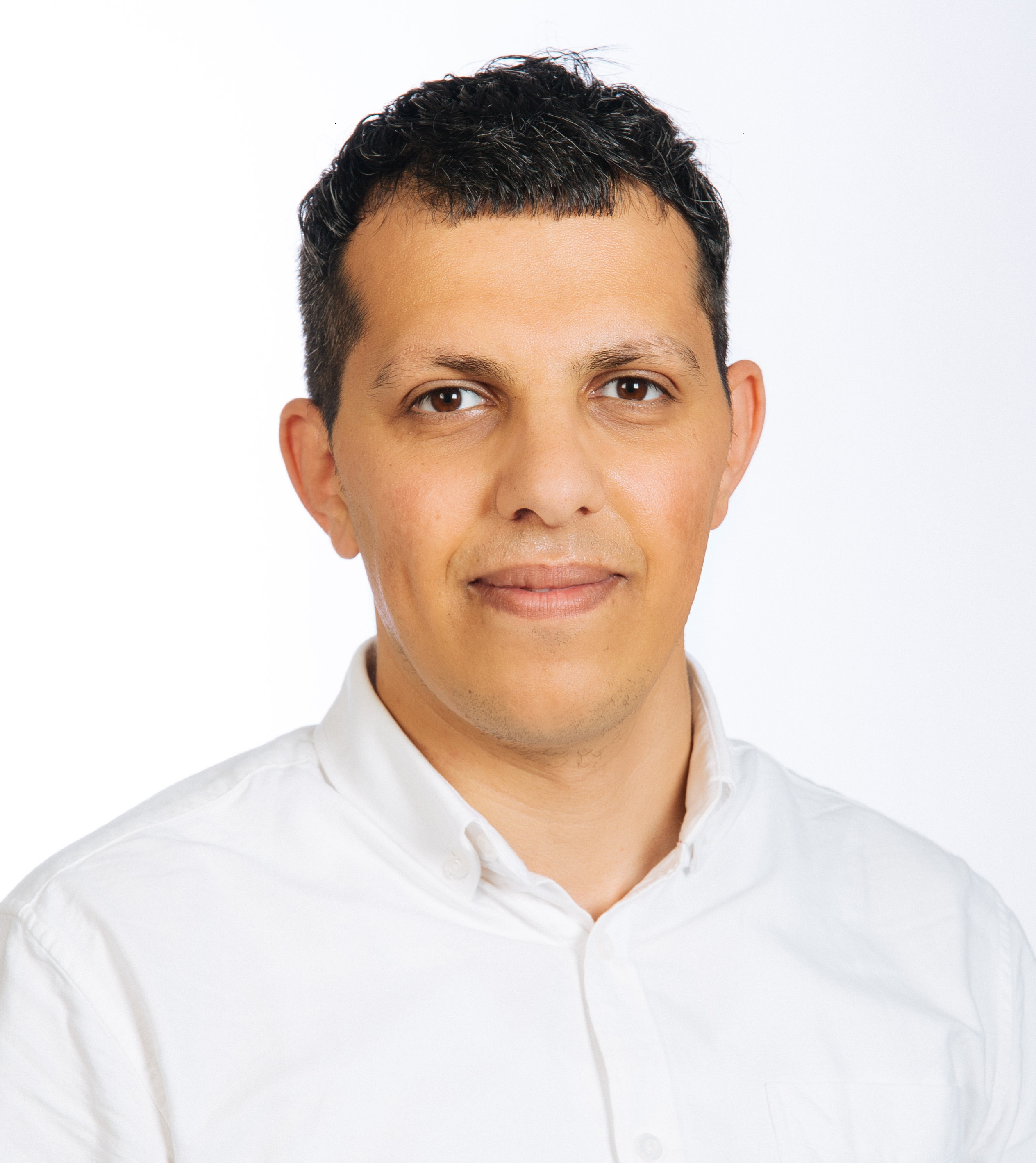}}]{Sahraoui Dhelim} is a senior researcher at University College Dublin, Ireland. He was a visiting researcher at Ulster University, UK (2020-2021). He obtained his PhD degree in Computer Science and Technology from the University of Science and Technology Beijing, China, in 2020. And a Master's degree in Networking and Distributed Systems from the University of Laghouat, Algeria, in 2014. And Bs degree in computer science from the University of Djelfa, in 2012. He serves as a guest editor in several reputable journals, including Electronics Journal and Applied Science Journal. His research interests include Social Computing, Smart Agriculture, Deep-learning, Recommendation Systems and Intelligent Transportation Systems.
\end{IEEEbiography}

\end{document}